\documentstyle[10pt,bezier]{article}
\begin{document}
\def \inbar{\vrule height1.5ex width.4pt depth0pt}
\def \xC{\relax\hbox{\kern.25em$\inbar\kern-.3em{\rm C}$}}
\def \xR{\relax{\rm I\kern-.18em R}}
\newcommand{\xZ}{Z \hspace{-.08in}Z}
\newcommand{\xbe}{\begin{equation}}
\newcommand{\xee}{\end{equation}}
\newcommand{\xbea}{\begin{eqnarray}}
\newcommand{\xeea}{\end{eqnarray}}
\newcommand{\xnn}{\nonumber}
\newcommand{\xkt}{\rangle}
\newcommand{\xbr}{\langle}
\newcommand{\xlll}{\left( }
\newcommand{\xrrr}{\right)}
\newcommand{\xcun}{\mbox{\footnotesize${\cal N}$}}
\title{Quantum Canonical Transformations and
Exact Solution of the Schr\"odinger Equation}
\author{Ali Mostafazadeh\thanks{E-mail: alimos@phys.ualberta.ca}\\ \\
Theoretical Physics Institute, University of Alberta, \\
Edmonton, Alberta,  Canada T6G 2J1.}
\date{ November 1996}
\maketitle

\begin{abstract}
Time-dependent unitary transformations are used to study the Schr\"odinger equation
for explicitly time-dependent Hamiltonians of the form $H(t)=\vec R(t)\cdot \vec J$, where
$\vec R$ is an arbitrary real vector-valued function of time and $\vec J$ is the angular momentum operator. The solution of the Schr\"odinger equation for the most general
Hamiltonian of this form is shown to be equivalent to the special case $\vec R=(1,0,\nu(t))$.
This corresponds to the problem of driven two-level atom for the spin half representation
of $\vec J$. It is also shown that by requiring the magnitude of $\vec R$ to depend on its direction in a particular way, one can solve the Schr\"odinger equation exactly. In
particular, it is shown that for every Hamiltonian of the form $H(t)=\vec R(t)\cdot \vec J$
there is another Hamiltonian with the same eigenstates for which the Schr\"odinger equation is exactly solved. The application of the results to the exact solution of the parallel transport equation and exact holonomy calculation for $SU(2)$ principal bundles (Yang-Mills gauge theory) is also pointed out. 
\end{abstract}
\vspace{.3cm}

\section{Introduction}

In non-relativistic quantum mechanics the dynamics of  pure states is determined by the Schr\"odinger equation,
	\xbe
	H\psi=i\dot\psi\;,
	\label{sch-eq}
	\xee
where $H$ is the Hamiltonian, $\psi$ is the state vector representing the state, 
the dot denotes a time-derivative, and $\hbar$ is set to unity.  In general the Hamiltonian
may be explicitly time-dependent, in which case the exact solution of the Schr\"odinger
equation is in general not known. In terms of the time-evolution operator $U=U(t)$ defined
by $U(t)\psi(0):=\psi(t)$, the Schr\"odinger equation (\ref{sch-eq}) is written as 
	\xbe
	H(t)U(t)=i \,\dot U(t)\,,~~~~~~~U(0)=1\;.
	\label{sch-eq-u}
	\xee
An alternative expression for this equation is $U(t)={\cal T}\exp [-i\int_0^t H(t')dt']$, where
${\cal T}$ denotes the time-ordering operator. The purpose of this article is to
derive some general sufficiency conditions to obtain the exact solution of
Eq.~(\ref{sch-eq-u}) for the dipole Hamiltonians:
	\xbe
	H(t)=\sum_{a=1}^3 R^a (t)J_a=\vec R\cdot\vec J\;,
	\label{h}
	\xee
where  $R^a$ are real functions of time which do not simultaneously vanish,
and $J_a$ are generators of the group $SU(2)$ in some irreducible representation. 

Following the same line of reasoning as in the Hamilton-Jacobi theory of classical mechanics, one can view the inverse $U^{-1}(t)=U^\dagger(t)$ of the evolution operator $U(t)$ as a time-dependent  quantum canonical transformation which sets the Hamiltonian to zero.  In order to see this more clearly, let us first recall that in quantum mechanics the role of  canonical transformations is played by unitary transformations of the Hilbert space. Now consider an arbitrary time-dependent unitary transformation $\psi(t)\to\psi'(t)={\cal U}(t)\psi(t)$. Requiring this transformation to preserve the form of the Schr\"odinger equation (\ref{sch-eq}), one has
	\xbea
	H(t)\rightarrow H'(t)&=&
	{\cal U}(t)H(t)\,{\cal U}^\dagger(t)-i\,{\cal U}(t)\,\dot{\cal U}^\dagger(t)\;,
	\label{trans}\\
	U(t)\rightarrow U'(t)&=&{\cal U}(t)U(t)\,{\cal U}^\dagger(0)\;.
	\label{trans-u}
	\xeea
Hence, $U^\dagger(t)$ induces a particular quantum canonical transformation which renders the transformed Hamiltonian zero.  In other words, if one views the effect of a quantum canonical transformation as a change of frame in the Hilbert space,  then the transformation induced by $U^\dagger (t)$ takes one to a moving frame in which the state vector is stationary, i.e., $\psi'(t)=\psi'(0)$. 

As it is manifestly seen from (\ref{trans}), quantum canonical transformations also resemble the non-Abelian gauge transformations of  particle physics. Therefore, in a sense solving the Schr\"odinger equation (\ref{sch-eq-u})  is equivalent to finding an appropriate gauge in which the state vector is stationary.\footnote{Note however that here there does not exist an analog of a non-Abelian gauge symmetry unless one deals with peculiar constraint systems such as those encountered in quantum cosmology.}

In this paper I shall try to demonstrate the utility of this simple observation in solving the Schr\"odinger equation for a large class of Hamiltonians of the form (\ref{h}). The basic idea
pursued in this paper is to find a series of  unitary (gauge) transformations which simplify the form of the Hamiltonian and yield previously unknown exactly solvable cases. Of course a priori  there is no systematic method of choosing appropriate gauge transformations. However it turns out that at least for the systems considered here, one is guided by basic group theoretical properties of angular momentum operators and methods of quantum adiabatic approximation.
Probably the most notable feature of this method which makes it so effective is its non-perturbative nature. 

\section{Dipole Hamiltonian and Parallel Transportation in $SU(2)$ Bundles}

Consider the Dipole Hamiltonian (\ref{h})
	\xbe
	H=H[R]=\vec R \cdot \vec J=r(\sin\theta\cos\varphi J_1+
	\sin\theta\sin\varphi J_2+\cos\theta J_3)=W(\theta,\varphi)J_3\,W^\dagger
	(\theta,\varphi)\,,
	\label{H}
	\xee
which describes the dynamics of a magnetic dipole in a changing magnetic field. Here
$\vec R:=(R^1, R^2,R^3)=(r,\theta,\varphi)$ corresponds to the magnetic field vector
expressed in units in which the Larmor frequency is set to unity, $(r,\theta,\varphi)$ are
spherical coordinates, and
	\xbe
	W(\theta,\varphi):=e^{-i\varphi J_3}
	e^{-i\theta J_2}e^{i\varphi J_3}\;.
	\label{w}
	\end{equation}
Then an arbitrarily changing magnetic field corresponds to a curve $C:[0,T]\to \xR^3$,
$\vec R=\vec R(t)=C(t)$. 

An application of the dipole Hamiltonian (\ref{H}) is in the parallel transportation in $SU(2)$ principal fiber bundles (Yang-Mills theory). This is easily seen by recalling that
parallel transportation \cite{nakahara} is defined in terms of a Lie algebra-valued one-form
(gauge potential) $A=A^a_\mu J_a dx^\mu$ according to
	\xbe
	g[{\cal C}]={\cal P}e^{-i\int_{\cal  C}A}=
	{\cal P}e^{-i\int_{\cal  C(0)}^{\cal C(T)}A^a_\mu J_a dx^\mu}=
	{\cal T}e^{-i\int_0^T  \dot x^\mu(t) A^a_\mu[x(t)] J_a }\;,
	\label{pa-tr}
	\xee
where ${\cal C}:[0,T]\to M$ is a curve in the base manifold $M$ of the bundle (spacetime in Yang-Mills theory), and $t$ is an arbitrarily chosen parameter of  the curve ${\cal C}$.
It is very easy to recognize the last expression on the right hand side of (\ref{pa-tr}) as 
the time-evolution operator $U(T)$ for a Hamiltonian of the form (\ref{H}) with $R^a=
\dot x^\mu(t) A^a_\mu[x(t)]$. Thus an exact solution of the Schr\"odinger equation for (\ref{H}) yields as a special case the solution for the problem of parallel transportation and in particular the calculation of the holonomy elements and Wilson loop integrals in Yang-Mills theory.

Let us next recall the basic properties of the most general Hamiltonians of the form (\ref{H}),
\cite{bohm-qm}. It is not difficult to see that the eigenvalues $E_n[R]$ and eigenvectors $|n;R\xkt$ of  $H[R]$ are given by:
	\xbea
	E_n[R]&=&E_n(r,\theta,\varphi)=E_n(r,0,0)=nr\:,~~~{\rm with}~~
	n=-j,-j+1,\cdots,j\;,
	\label{eg-va}\\
	|n;R\xkt&=&|n;(r,\theta,\varphi)\rangle=|n;(r_0,\theta,\varphi)\rangle=
	W(\theta,\varphi)
	|n;(r_0,0,0)\rangle\;, ~~~\theta\in[0,\pi),~\varphi\in[0,2\pi),
	\label{eg-ve}
	\xeea
where $j$ corresponds to the spin $j$-representation of $SU(2)$ and determines the Hilbert space, and $(r_0,\theta_0,\varphi_0):=(r(0),\theta(0),\varphi(0))$. Hence, the Hamiltonian
$H$ is non-degenerate for $r\neq 0$. In order to avoid the complications caused by the
sudden collapse of all the energy eigenvalues which occurs at $r=0$, I shall only consider
the case where the curve $C$ does not pass through the origin, i.e., $C(t)=\vec R(t)\in
\xR^3-\{ 0\}$.	
	
Note that $\{|n;(t,\theta,\varphi)\xkt\}$ forms a single-valued orthonormal basis of the Hilbert space for all $\varphi\in[0,2\pi)$ and $\theta\in[0,\pi)$ and that $|n;(r,0,0)\rangle$ are the eigenvectors of  $H(r,\theta=0,\varphi=0)=rJ_3$, i.e., $J^3|n;(r,0,0)\rangle= n|n;(r,0,0)\rangle$.
For $\theta=\pi$, $|n;(t,\theta,\varphi)\xkt$ are not single-valued. This is due to the fact that
the spectral bundle over $\xR^3-\{0\}$, \cite{simon}, which yields $|n;(t,\theta,\varphi)\xkt$
as its local basis sections is not trivial. In the parameterization of $\xR^3-\{0\}$ used here
the negative $z$-axis ($\theta=\pi$) is not included in the patch over which
$|n;(t,\theta,\varphi)\xkt$ are well-defined. To treat the negative $z$-axis, one must switch to
new coordinates $R':=(r'=r, \theta'=\pi-\theta,\varphi'=\varphi)$. The eigenvectors $|n;R'\xkt$
will then be single-valued everywhere except on the positive $z$-axis. In the following, I shall assume for simplicity but without loss of generality that the curve $C$ does not intersect the negative $z$-axis. In the general case where $C$ intersects the negative $z$-axis, one must make appropriate $U(1)$ gauge transformations which relate $|n;R\xkt$ and $|n;R'\xkt$, \cite{bohm-qm}.

I shall also assume that $|n;(0,0)\xkt$ and therefore $|n;R\xkt$ are eigenvectors of the total angular momentum operator, i.e., the Casimir operator $|\vec J|^2=\sum_{a=1}^3 J^2_a$.
This is always possible unless $R^a$ are also quantized, \cite{bohm-qm}. The latter case
will not be considered in the present article.

\section{Adiabatic Approximation and Reduction to Two-Dimensions}

In order to implement the idea of successive quantum canonical transformations, I shall begin
using the results of the adiabatic approximation.  One knows from the standard arguments of
Born and Fock \cite{fock} and Kato \cite{kato}, that if the time-dependence of the Hamiltonian is adiabatic, then in time the eigenstates of the initial Hamiltonian $H[R(0)]$ evolve into the eigenstates of the  Hamiltonian $H[R(t)]$. This is actually very easy to see if one differentiates both sides of the eigenvalue equation
	\xbe
	H(t)|n;t\xkt=E_n(t)|n,t\xkt\;,
	\label{eg-va-eq}
	\xee
and computes the inner product of both sides of the resulting equation with $|m;t\xkt$ for some $m\neq n$. This yields
	\xbe
	A_{mn}:=\xbr m;t|\frac{d}{dt}|	n;t\xkt=\frac{\xbr m;t|
	\dot H(t)|n;t\xkt}{E_n(t)-E_m(t)}\;,~~~~~m\neq n\;.
	\label{ad-eq}
	\xee
In Eqs.~(\ref{eg-va-eq}) and (\ref{ad-eq}),  $H(t):=H[R(t)],~|n;t\xkt:=|n;R(t)\xkt$, and $E_n(t):=E_n[R(t)]$. The adiabatic approximation is valid if and only if the right hand side of (\ref{ad-eq}) is negligible.  Now let us choose $\psi(0)=|n;0\xkt$, then in view of (\ref{ad-eq}), it is easy to show that $\psi(t)=e^{i\alpha_n(t)}|n;t\xkt$ does solve the schr\"odinger equation provided that
	\xbe
	\alpha_n(t):=\delta_n(t)+\gamma_n(t)\;,~~~~
	\delta_n(t):=-\int_0^t E_n(t')dt'\,~~~~
	\gamma_n(t):=i\int_0^t A_{nn}(t')dt'\,.
	\label{adg}
	\xee
The phase angles $\alpha_n(T)$, $\delta_n(T)$, and $\gamma_n(T)$ for a closed curve $C$ are known as the total, dynamical, and adiabatic geometrical (Berry) phase angles, \cite{berry1984}. 

The adiabatic approximation which also includes the geometric phase effects,
corresponds to approximating the time-evolution operator $U(t)$ with
	\xbe
	U_0(t):=\sum_ne^{i\alpha_n}|n;t\xkt\xbr n;0|\;.
	\label{uo}
	\xee
In general the approximation $U\approx U_0$ is not valid. However, one can
compute $U_0$ in terms of the eigenvalues and eigenvectors of the Hamiltonian
and use $U_0^\dagger$ to perform a quantum canonical transformation. In the remainder
of this section I shall show that indeed this canonical transformation simplifies the form of the Hamiltonian considerably.

In order to do this one must first calculate the matrix elements $A_{mn}$ which enter the calculation of $\alpha_n$ and especially the term ${\cal U}\dot{\cal U}^\dagger$ in
Eq.~(\ref{trans}) with ${\cal U}=U_0^\dagger$. This rather lengthy calculation leads to
	\xbea
	A_{mn}&=&i\left[ m(1-\cos\theta)\delta_{mn}+
	\frac{1}{2}\: \sin\theta\: (e^{i\varphi}C_m\delta_{m\,n-1}+
	e^{-i\varphi}C_n\delta_{m-1\,n})\right]\:\dot\varphi+\xnn\\
	&&\frac{1}{2}\:(e^{i\varphi}C_m\delta_{m\,n-1}-e^{-i\varphi}C_n\delta_{m-1\,n})
	\:\dot\theta\;,
	\label{amn}
	\xeea
where $C_m:=\sqrt{(j-m)(j+m+1)}=C_{-m-1}$, and extensive use is made of the properties of $J_a$ and $J_\pm:=J_1\pm i J_2$, particularly 
	\xbea
	e^{-i\beta J_a}J_b \:e^{i\beta J_a}&=&\cos\beta\,J_b+\epsilon_{abc}
	\sin\beta\, J_c\;,~~~~~ a\neq b\;,\xnn\\
	J_\pm\,|m;(r,0,0)\rangle&=&\hbar\,C_{\pm m}\:|m\pm 1;(r,0,0)\rangle\;,\xnn
	\xeea
where $\epsilon_{abc}$ are components of the totally anti-symmetric Levi Civita symbol, with
$\epsilon_{123}=1$. Furthermore, one can easily show that $\alpha_n=n\alpha,~
\delta_n=n\delta$, $\gamma_n=n\gamma$, where $\alpha=\delta+\gamma$, and
	\xbe
	\delta=-\int_0^t r(t')\:dt'\;,~~~~~\gamma=-\int_0^t[1-\cos\theta(t')]\,\dot\varphi(t')\:dt'\;.
	\label{delta-gamma}
	\xee
These relations are then used to write down the expression for $U_0$, namely,
	\xbe
	U_0(t)=W(\theta(t),\varphi(t))\,e^{i\alpha(t) J_3}W^\dagger(\theta_0,\varphi_0)\;,
	\label{u1=}
	\xee
where $W$ is defined in Eq.~(\ref{w}).

Next let us set ${\cal U}=U_0^\dagger$ in Eq.~(\ref{trans}). Then using Eq.~(\ref{amn}), one finds
the expression for the transformed Hamiltonian
	\xbe
	H_0(t)= \frac{1}{2}\: W(\theta_0,\varphi_0) \left[ 
	\Omega(t)\,J_+ + \Omega^*(t)\,J_-\right] W^\dagger
	(\theta_0,\varphi_0)\;,
	\label{H1}
	\xee
where
	\xbe
	\Omega(t):=e^{-i[\alpha(t)+\varphi(t)]}\:[\sin\theta(t)\:
	\dot\varphi(t)+i\dot\theta(t)]\;.
	\label{Omega}
	\xee

One can easily see that if $\theta_0=\varphi_0=0$, then $W(\theta_0,\varphi_0)=1$ and the expression (\ref{H1}) for the transformed Hamiltonian simplifies considerably. Hence, it is convenient to choose the coordinate system in such a way that $\theta_0=\varphi_0=0$, i.e., $\vec R(0)=(0,0,r_0)$, or alternatively make a further constant unitary transformation using
${\cal U}=W^\dagger(\theta_0,\varphi_0)$ which leads to the Hamiltonian
	\xbe
	H_1(t)=\omega(t)\left[ \cos\sigma(t)J_1-\sin\sigma(t)J_2\right]\;,
	\label{H1'}
	\xee
where $\Omega=:\omega e^{i\sigma}$, i.e.,
	\xbea
	\omega(t)&:=&\sqrt{\dot\theta^2+\sin^2\theta\,\dot\varphi^2}\;,~~~~
	\sigma(t)\::=\:-\alpha-\varphi+\xi~~~~~~{\rm mod}~2\pi\,,\xnn\\
	\cos\xi&:=&\frac{\sin\theta\:\dot\varphi}{\omega}\,,~~~~
	\sin\xi\::=\:\frac{\dot\theta}{\omega}\,\xnn
	\xeea
One can also combine the two unitary transformations by transforming $H$ by ${\cal U}=U_1^\dagger(t)$ with $U_1(t):=U_0(t)W(\theta_0,\varphi_0)$.

The Hamiltonian (\ref{H1'}) describes the dynamics of a magnetic dipole in a
time-dependent magnetic field which is confined to the $x$-$y$ plane, i.e., a
Hamiltonian of the form (\ref{H})  corresponding to a planar curve $C_1:[0,T]\to \xR^2-\{0\}$. Hence,  the canonical transformation induced by $U_1$ reduces the three-dimensional
problem to a two-dimensional one. 

\section{Exactly Solvable Cases}

Consider the Schr\"odinger equation for the Hamiltonian $H_1$. If  the angular variable $\sigma$ happens to be constant, then this equation  can be easily integrated. This is simply because in this case $H_1$ at different times commute and the transformed evolution operator is obtained by its exponentiation, i.e.,
	\xbe
	U'(t)=e^{-i\ell(t)[\cos\sigma_0J_1-\sin\sigma_0J_2]}\;,
	\label{u'}
	\xee
where $\ell(t):=\int_0^t\omega(t')dt'$,
	\xbe
	\sigma_0:=\sigma(0)=-\varphi_0+\xi(0)=
	-\varphi_0+tan^{-1}[ \frac{\dot\theta(0)}{\sin\theta(0)\dot\varphi(0)}]
	=-\varphi_0+ tan^{-1}[\frac{\theta'(\varphi_0)}{\sin\theta(\varphi_0)}]\;,
	\xee
and $\theta':=d\theta/d\varphi$. 

Having found the evolution operator $U'$ for $H_1$, one can use Eq.~(\ref{trans-u})
to write down the solution of the original Schr\"odinger equation (\ref{sch-eq-u}). This
yields
	\xbe
	U(t)=U_1(t)U'(t)U_1^\dagger(0)=U_0(t)W(\theta_0,\varphi_0)U'(t)
	W^\dagger(\theta_0,\varphi_0), ~~~~{\rm for}~~~~\sigma(t)=\sigma_0\;.
	\label{exact}
	\xee
Note that the parameters $\sigma_0$ and $\ell$ which enter the expression for $U(t)$
are geometric quantities associated with the projection $C'$ of the curve $C$ onto the unit sphere centered at the origin. In particular,  $\ell$ is the length of $C'$. Furthermore  for those portions of the curve $C$ which project to a single point for an extended period of time, $\omega$ and consequently $H_1$ vanish. This is reminiscent of the known fact  that the adiabatic approximation is exact when  the eigenvectors of the Hamiltonian are stationary. 

Another way of arriving at the same conclusion is by performing another quantum canonical transformation with ${\cal U}=U_2^\dagger:=e^{-i\sigma(t)J_3}$. This leads to the transformed Hamiltonian
	\xbe
	H_2=\omega(t)J_1+\dot\sigma(t)J_3\;.
	\label{H2}
	\xee
Clearly for $\sigma=$ const.\ the Schr\"odinger equation for $H_2$ is exactly solvable. Making a further canonical transformation with ${\cal U}=U_3^\dagger:=e^{i\ell(t)J_1}$, one obtains
	\xbe
	H_3=\dot\sigma[\cos\ell(t)\,J_3+\sin\ell(t)\,J_2]\;.
	\label{H3}
	\xee
which vanishes identically for $\sigma=$ const. Therefore, as expected the combined transformation ${\cal U}=(U_1 U_2 U_3)^\dagger$ leads to a frame in which the Hamiltonian
vanishes and the state vector is stationary. Hence, the original time-evolution operator is given by $U=U_1 U_2 U_3$.

Let us next re-express the condition $\sigma=$ const.~in terms of the original variables. Requiring
$\dot\sigma=0$, one finds the equivalent condition: $r(t)=r_*(t)$, where
	\xbe
	r_*(t):=\cos\theta\:\dot\varphi-
	\frac{  \frac{d}{dt}(\frac{\dot\theta}{\sin\theta\,\dot\varphi})}{
	1+(\frac{\dot\theta}{\sin\theta\,\dot\varphi})^2}=\left[\cos\theta-
	\frac{\frac{d}{d\varphi}(\frac{\theta'}{\sin\theta})}{
	1+(\frac{\theta'}{\sin\theta})^2}\right]\:\dot\varphi\;.
	\label{r=}
	\xee
Therefore, one has:
	\begin{itemize}
	\item[] {\bf Lemma~1:} {\em The exact solution of the Schr\"odinger equation
	(\ref{sch-eq-u}) is given by (\ref{exact}) provided that the magnitude of the
	magnetic field depends on its direction according to $r(t)=r_*(t)$.}
	\end{itemize}
This is quite remarkable, for it indicates that for every Hamiltonian of the form (\ref{H}) for which $r_*$ does not vanish for extended periods of time, 
there exists another Hamiltonian with the same eigenvectors\footnote{Note that the eigenvectors only depend on the direction of the magnetic field.}  whose Schr\"odinger equation is exactly solvable. Note that for time intervals during which $r_*<0$, one can consider the time-reversed system where $r_*>0$. The evolution operator obtained for the time-reversed system yields the original time-evolution operator upon inversion. This leaves only the cases where $r_*$ vanishes, i.e., either $\dot\varphi=0$ or $\theta'=\sin\theta\tan[\sin\theta+c]$ for some constant $c$. A simple case where the latter equation is satisfied is $\theta=\pi/2$ and $c=-1$. This means that for the planar curves with $\theta=\pi/2$ such as $C_1$, one cannot enforce the condition $r=r_*$ and the exact solution cannot be obtained in this way. Therefore a direct
repetition of the same procedure for the Hamiltonian $H_1$ will not lead to the exact solution.
In the remainder of this section I shall demonstrate, however, that by a straightforward redefinition of the time one can generalize Lemma~1 further.

Let us first note that for the case where $\omega=0$ the exact solution is given by the adiabatic approximation. Hence, without loss of generality one can restrict to the case $\omega\neq 0$. In this case the length $\ell$ of the projection $C'$ of the curve $C$ is a monotonically increasing function of time $t$. Therefore it can be used to parameterize the evolution of the system, i.e., replace $t$. Changing variables from $t$ to $\ell$ in the Schr\"odinger equation for the Hamiltonian $H_1$ and making use of $\omega\neq 0$, one has
	\xbe
	\bar H_1(\ell)\bar U_1(\ell)=i\frac{d}{d\ell}\:\bar U_1(\ell)\;,
	\label{sch-eq-2l}
	\xee
where
	\xbe
	\bar H_1(\ell):=\cos\sigma(\ell)J_1-\sin\sigma(\ell)J_2=
	e^{i\sigma(\ell) J_3}J_1e^{-i\sigma(\ell) J_3}\;.
	\label{H1l}
	\xee
This reduces the problem to the case of a magnetic field which traces a circular path in the $x$-$y$ plane with an angular frequency, $\nu:=d\sigma/d\ell=(r-r_*)/\omega$. Note that the
presence of $\omega(t)$ on the right hand of Eq.~(\ref{H1'}) is quite essential in the redefinition
of time.

Let us next transform to the rotating frame defined by ${\cal U}=\bar U_2^\dagger(\ell):=e^{-i\sigma(\ell) J_3}$. In view of Eq.~(\ref{trans}), this leads to the transformed Hamiltonian
	\xbe	
	\bar H_2(\ell)=J_1+\nu(\ell)\, J_3\;,
	\label{H2l}
	\xee
which describes a  magnetic field with a constant $x$-component and a variable $z$-component. Such systems are widely encountered in the study of nuclear and optical magnetic resonance. For a recent study of an iterative solution of the Schr\"odinger equation for this Hamiltonian see Refs.~\cite{mu,ro}.

Note that for $\nu=\nu_0=$ const.,  $\bar H_2$ is constant. Hence, the transformed time-evolution operator is given by $\bar U_3(\ell):=\exp[-i\ell(J_1+\nu_0 J_3)]$, and
one has
	\xbe
	U(t)=U_1(t)\bar U_2(\ell(t))\bar U_3(\ell(t))\bar U_2^\dagger(0)U_1^\dagger(0)=
	U_0(t)W(\theta_0,\varphi_0)\bar U_2(\ell(t))\bar U_3(\ell(t))\bar U_2^\dagger(0)
	W^\dagger(\theta_0,\varphi_0)\;.
	\label{exact2}
	\xee
This concludes the derivation of the exact solution of the schr\"odinger equation for the case where $\sigma(\ell)=\sigma_0+\nu_0 \ell$, alternatively, $r(t)=r_*(t)+\nu_0\,\omega(t)$. This
is a generalization of Lemma~1. It states that even for the time periods during which $r_*=0$, the above procedure still leads to exactly solvable Schr\"odinger equations. More precisely, the following lemma holds.
	\begin{itemize}
	\item[] {\bf Lemma~2:} {\em The exact solution of the Schr\"odinger equation
	(\ref{sch-eq-u}) is given by Eq.~(\ref{exact2}), provided that the magnitude of the
	magnetic field depends on its direction according to $r(t)=r_*(t)+\nu_0\omega(t)$, 
	for some constant $\nu_0$.}
	\end{itemize}

A direct consequence of this result is
	\begin{itemize}
	\item[] {\bf Corollary:} {\em For every Hamiltonian of the form~(\ref{H}), there
	exists another Hamiltonian with the same eigenvectors for which the Schr\"odinger
	equation is exactly solvable.}
	\end{itemize}

\section{Conclusion}

In this paper I have used a variety of time-dependent unitary transformations of the Hilbert space to obtain  the exact solution of the schr\"odinger equation for a large class of explicitly time-dependent dipole Hamiltonians. This involved redefinition of the time variable which
was a consequence of transforming to a moving frame via the inverse of the adiabatically approximate time-evolution operator. In this frame the natural choice for
the evolution parameter turned out to be the length of the projection of the curve $C$ traced by the tip of the magnetic field onto the unit sphere centered at the origin.

The reduction of the general problem to that of  the Hamiltonian  $\bar H_2=J_1+\nu(\ell)J_3$ may also be used to set up an approximation scheme for  large $\omega$. This is due to the fact
that $\nu=(r-r_*)/\omega$ may be neglected for large $\omega$, in which case Lemma~2
provides the solution. This is particularly effective for the dipole Hamiltonians which
correspond to a planar curve $C$, for which $r_*=0$, e.g., $\bar H_1$. For these
Hamiltonians,  the approximation is valid if  the parameter $r(t)/\omega(t)$ is negligible.
Note also that for such Hamiltonians if $r(t)$ and $\omega(t)$ are proportional, then
Lemma~2 yields the exact solution to the Schr\"odinger equation.

Moreover, by successive application of the method used in this
reduction, i.e., by replacing the original Hamiltonian $H$ by $\bar H_1$ and repeating
the same analysis, one obtains an iterative  solution of the Schr\"odinger equation which yields a product expansion of the time-evolution operator. The condition of the termination of this expansion after a finite number of iterations may seem to lead to (possibly) more general
exactly solvable cases. It turns out that this is in fact not the case.
This is because enforcing the condition that the above expansion be terminated after
the second iteration leads to $\omega=$ const.~, which is certainly not more general
than the conditions  of Lemma~1 and Lemma~2.  This marks a unique property of
the Hamiltonians of type $H(t)=J_1+R^3(t)J_3$.

The results of this paper have direct applications in the computation of the holonomy elements and Wilson loop integrals in Yang-Mills theory where the gauge group is $SU(2)$.\footnote{Clearly the $U(2)$ case can also be handled similarly.} In this case
the original parameters $R^a$ of the Hamiltonian (\ref{H}) are identified with $\dot x^\mu A_\mu^a$, where $(A_\mu^a)$ corresponds to the local connection one-form (gauge potential)
and the gauge transformations correspond to quantum canonical transformations of the associated Hamiltonian. Another area of application of the results of this article is in the calculation of non-Abelian $U(2)$ geometric phases \cite{wz} such as those encountered in the
study of the three-level systems, \cite{p20}.

\section*{Acknowledgements}
I would like to thank Dr.~M.~Razavi for invaluable comments and suggestions and acknowledge the financial support of the Killam Foundation of Canada.

\end{document}